\documentclass{article}
\usepackage{ijcai25}

\usepackage{times}
\usepackage{soul}
\usepackage{url}
\usepackage[hidelinks]{hyperref}
\usepackage[utf8]{inputenc}
\usepackage[small]{caption}
\usepackage{graphicx}
\usepackage{multirow}
\usepackage{pifont}   
\usepackage{amsmath}
\usepackage{amsthm}
\usepackage{booktabs}
\usepackage{algorithm}
\usepackage{algorithmic}
\usepackage[table]{xcolor}
\definecolor{lightgray}{gray}{0.9}
\usepackage[switch]{lineno}

\title{See What I Mean? CUE: A Cognitive Model of Understanding Explanations}

\author{
Tobias Labarta$^1$\and
Nhi Hoang$^1$\and
Katharina Weitz$^1$\and\\
Wojciech Samek$^{1,2,3,\dagger}$\and
Sebastian Lapuschkin$^{1,4,\dagger}$\and
Leander Weber$^{1,\dagger}$
\affiliations
$^1$ Fraunhofer Heinrich Hertz Institute, Berlin, Germany\\
$^2$ Technische Universität Berlin, Berlin, Germany\\
$^3$ BIFOLD – Berlin Institute for the Foundations of Learning and Data, Berlin, Germany\\
$^4$ Centre of eXplainable Artificial Intelligence, Technological University Dublin, Dublin, Ireland\\
\emails{\texttt{$^\dagger$\{wojciech.samek,sebastian.lapuschkin,leander.weber\}@hhi.fraunhofer.de}}
}

\begin{document}

\maketitle

\begin{abstract}
As machine learning systems increasingly inform critical decisions, the need for human-understandable explanations grows. Current evaluations of Explainable AI (XAI) often prioritize technical fidelity over cognitive accessibility which critically affects users, in particular those with visual impairments. We propose CUE, a model for \textbf{C}ognitive \textbf{U}nderstanding of \textbf{E}xplanations, linking explanation properties to cognitive sub-processes: legibility (perception), readability (comprehension), and interpretability (interpretation). In a study (\textit{N}=455) testing heatmaps with varying colormaps (BWR, Cividis, Coolwarm), we found comparable task performance but lower confidence/effort for visually impaired users. Unlike expected, these gaps were not mitigated and sometimes worsened by accessibility-focused color maps like Cividis. These results challenge assumptions about perceptual optimization and support the need for adaptive XAI interfaces. They also validate CUE by demonstrating that altering explanation legibility affects understandability. We contribute: (1) a formalized cognitive model for explanation understanding, (2) an integrated definition of human-centered explanation properties, and (3) empirical evidence motivating accessible, user-tailored XAI.
\end{abstract}

\section{Introduction}

Machine Learning (ML) systems are increasingly used to support high-stakes decisions. Consequently, the need for human-understandable explanations has become a central concern in the field of Explainable AI (XAI) \cite{doshi2017towards,lage_evaluation_2019,arrieta2020explainable,longo2024explainable}. However, most current evaluation methods focus on technical properties such as faithfulness \cite{samek_evaluating_2017,agarwal2020explaining}, robustness \cite{montavon2018methods,bhatt2020evaluating}, or semantical properties such as polysemanticity or clarity \cite{dreyer2024pure,dreyer2025mechanistic}. Although these methods consider crucial aspects of XAI evaluation, they are not addressing visual properties, impacting how well users can perceive, process, and make sense of the explanations presented \cite{liao2021human,donoso2023towards}. This challenge becomes especially critical as explanations should be accessible to a diverse spectrum of users, including those with visual impairments and other cognitive disabilities \cite{nwokoye2024survey,tielman2024explainable}. Given this disconnect, we focus in this work on understanding \textit{how} we understand visual explanations.

In the context of our work, we define \textit{visual explanations} as any information presented to users with the intention of explaining a machine learning model’s inference, provided that it can be perceived via the human visual system. This includes images, figures, and text, excluding non-visual modalities (e.g. audio) \cite{becker2024audiomnist}.

To better understand how users engage with visual explanations, we turn to the concept of readability, originally developed in linguistics to describe “what makes some texts easier to read than others” \cite{dubay_principles_2004}. Foundational work  \cite{legge_psychophysics_1985,legge_psychophysics_1987} showed how visual factors like character size and contrast affect reading speed, while others \cite{dale_concept_1949} framed readability as factors affecting readers’ success with printed material.

A crucial distinction has emerged in linguistics between \textit{legibility}, the visual clarity of text (e.g., font size, layout, contrast), and \textit{readability}, the cognitive ease of understanding what is read \cite{tekfi_readability_1987,hossain_concept_2024}. A document with large font size may be visually legible but difficult to read due to dense or overly technical language. These insights have extended beyond text to domains such as code comprehension, interface design, and data visualization \cite{zuffi_human_2007,strizver_type_2013,oliveira_evaluating_2020}. 

As the research of data visualization has evolved, a field concerned with optimizing visualization of complex datasets for a broad audience \cite{brodlie2012review}, the study of readability and legibility has expanded beyond linguistics to the realm of graphical interpretation. In the field of data visualizations, legibility refers to the clarity of individual visual elements (such as bars or lines), while readability involves understanding the overall narrative conveyed by these elements \cite{franconeri_science_2021,cabouat_previs_2024}. Together, these concepts play a critical role in minimizing cognitive load and maximizing the effectiveness of a visualization by facilitating easier interpretation.

Explainable AI visualizations, such as heatmaps \cite{zhou2016learning} and concept visualizations \cite{achtibat2023attribution}, can be seen as a specialized form of data visualization. Similar to traditional data visualizations,  they rely on visual encodings to communicate complex information, but they introduce unique challenges related to interpreting model behavior. While there has been a growing body of research directed to \textit{designing} XAI, it has primarily focused on the general interface and interaction design \cite{wolf2019explainability,liao2020questioning,chromik2021human,mohseni2021multidisciplinary}. We suggest that findings from data visualization research, particularly those not only concerning with optimizing layout and interaction design but also with specific visual attributes such as color intensity, contrast, or brightness, offer valuable insights for improving the understanding of explanations.

With this work, we make the following contributions:
\begin{itemize}
    \item An integrated definition of human-centred explanation properties that determine explanation understandability: legibility, readability, interpretability --- each aligned with a distinct cognitive sub-process of understanding: perception, comprehension, and interpretation.
    \item We propose CUE, a formalized cognitive model of understanding explanations, that uniquely links the external properties of explanations to internal user cognitive processes. It thereby integrates core concepts of human-centred XAI into a unified framework.
    \item We apply and evaluate CUE in a user study focused on heatmap explanations. We observe a significant impact of colormap choices on visually impaired users' interaction with heatmap explanations, underscoring the effect of explanation properties on user understanding.
\end{itemize}

\section{Related Work}
Layer-wise Relevance Propagation (LRP) \cite{bach2015pixel}, Local Interpretable Model-agnostic Explanations (LIME) \cite{ribeiro_why_2016}, and Gradient-weighted Class Activation Mapping (Grad-CAM) \cite{selvaraju_grad-cam_2017} are common methods in XAI for compiling attributions. LRP and Grad-CAM are model-specific, using internal structures (e.g., gradients), while LIME is model-agnostic, relying on local perturbations \cite{speith_conceptualizing_2024}.

Heatmaps are an essential visualization technique in XAI, using color gradients to depict continuous, non-binary feature attribution, activation, attention, or saliency across activation tensors \cite{gu_complex_2022}. Formally defined as $n$-dimensional spatial projections of model-derived attribution scores ($n \geq 2$), heatmaps are commonly applied in the image domain to highlight which input features contribute to model predictions \cite{nauta_anecdotal_2023}. In detection tasks, they localize specific elements within complex images, while in classification, they help users assess whether predictions rely on consistent and task-relevant regions. Techniques for generating heatmaps include sensitivity-based methods, deconvolution, and LRP \cite{samek_evaluating_2017,selvaraju_grad-cam_2020,richard_deep_2021}.

While several studies have evaluated attribution techniques using heatmap visualizations, they have been criticized for insufficient consideration of cognitive processes \cite{lopes2022xai} or for limited generalizability due to a narrow focus on specific explanation techniques or evaluation methods \cite{kim2024human}. In contrast, \cite{kares2025makes} proposed a broader evaluation framework that integrates subjective, objective, and computational measures. 

Recent frameworks, such as PREVis \cite{cabouat_previs_2024}, have made strides toward formalizing understanding within the domain of data visualization. PREVis assesses understandability, layout clarity, and data value/pattern readability. These frameworks aim to assess and improve how well users can comprehend complex visualizations, offering valuable insights to improve the design of visual elements and their interpretation.

A recent contribution by \cite{speith_conceptualizing_2024} introduces an abilities-based framework for understanding in XAI, distinguishing six capacities: \textit{recognizing}, \textit{assessing}, \textit{predicting}, \textit{intervening}, \textit{explaining}, and \textit{designing}. This approach emphasizes tailoring explanations to context-specific stakeholder needs. In the given context, that could be applicants assessing fairness versus developers designing systems. Unlike traditional monolithic approaches to XAI, the framework integrates insights from philosophy, psychology, and computer science, linking user abilities to goals such as fairness and trust.

Whilst \cite{speith_conceptualizing_2024} emphasize the abilities users need to interact with AI explanations, our model focuses on the \textit{process} of understanding. We show how sensory perception (legibility) evolves into causal reasoning (interpretability) through structured stages. Speith provides a foundation for the actions users can take based on explanations, while our model outlines how users cognitively reach that point. Together, these frameworks offer complementary insights for XAI design, aligning ethical and functional goals with a cognitive pathway. 

\section{CUE: A formalized Cognitive Model of Understanding Explanations}

\begin{figure*}[t]
    \centering
    \includegraphics[width=0.8\textwidth]{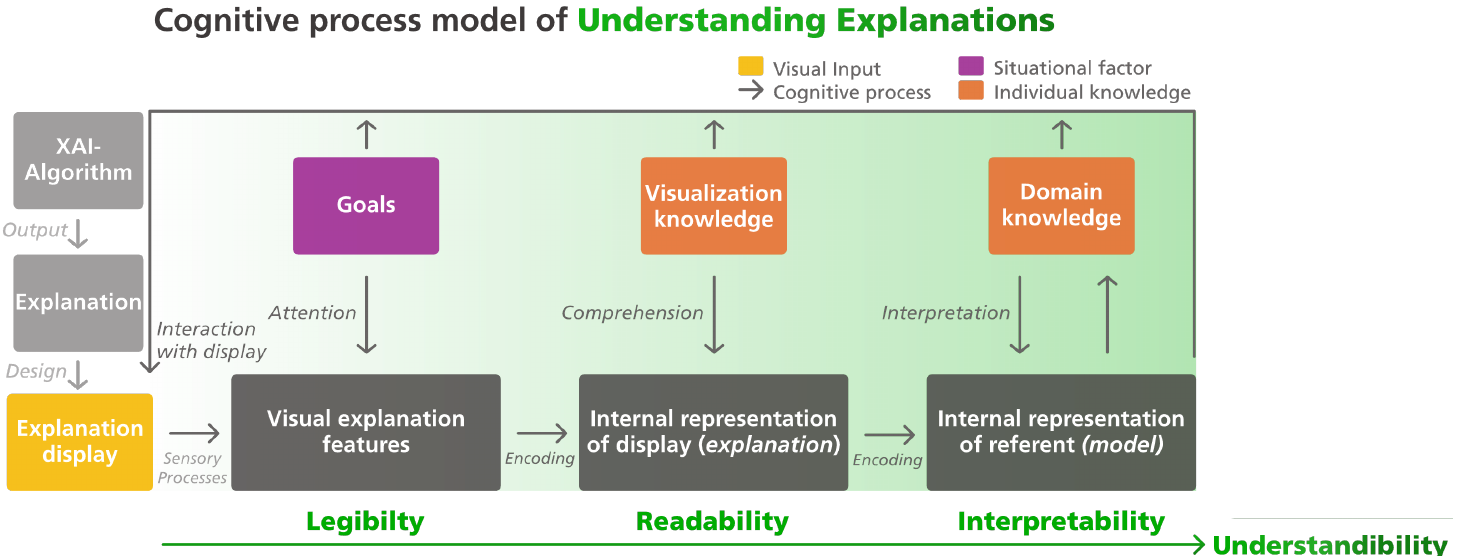}
    \caption{Adapted from the work of Cabouat et al. \protect\cite{cabouat_previs_2024}, we propose CUE, a \textit{cognitive model for understanding explanations} and integrate it with our definition of \textit{three properties of XAI understandability}: legibility, readability and interpretability.}
    \label{fig:cognitive_model}
\end{figure*}

According to \cite{cabouat_pondering_2023}, reading a visualization requires transforming raw visual features into meaningful internal representations, a process we extend to the understanding of explanations. Building on prior cognitive models of visualization comprehension \cite{cabouat_previs_2024}, we propose a structured framework for modeling the \textbf{C}ognitive \textbf{U}nderstanding of \textbf{E}xplanations --- \textbf{CUE}.

\subsection{The CUE Model in Detail} The cognitive model CUE considers both the individual elements that shape the interaction with explainable AI systems and the key processes that guide how explanations are understood. This model, as presented in Figure \ref{fig:cognitive_model}, frames explanation as a multi-step process combining the explanation itself, the way it is displayed, and how individuals engage with and interpret the information.

At the center of CUE is the XAI algorithm, which generates an output, the explanation, designed to make the ML model’s inference understandable to the user. Once created, the explanation is presented through an explanation display, which can take various forms such as verbal, visual, or diagrammatic representations of the explanation \cite{lim_diagrammatization_2023}. The choice of display plays a critical role in how effectively individuals can perceive and engage with the explanation \cite{narayanan_how_2018,dhanorkar_who_2021,poursabzi-sangdeh_manipulating_2021,kaur_sensible_2022}.

The individual engages with the explanation through sensory processes, motivated by explanation goals. These goals define what the individual seeks to achieve, such as understanding \textit{why} a particular prediction was made or \textit{how} to improve the model’s performance. Attention is directed toward the relevant visual explanation features, whose ease of perception is defined as legibility. The perceived explanation features are then encoded into an internal representation of the explanation, a mental model that helps the individual make sense of the information. This step of comprehension confines the readability of explanations. Crucial to this process is the individual's level of visualization knowledge, i.e. their familiarity with the conventions used in the explanation, such as heatmaps or conceptual visualizations. 

Finally, their domain knowledge, or expertise in the subject area, allows them to interpret the explanation and connect it to the underlying model, the last step of understanding explanations.

\paragraph{Understandability determines Understanding.} A central idea of the cognitive model for understanding explanations is that \textbf{understandability is an inherent property of the explanation}. This property reflects the explanation’s capacity to support the user’s overall \textbf{cognitive process of understanding}. Crucially, this cognitive process is not monolithic, but consists of three interdependent sub-processes: perception, comprehension, and interpretation. These sub-processes are supported, respectively, by three sub-properties of explanation understandability: legibility, readability, and interpretability. While these properties are observable and designable qualities of the explanation, the associated cognitive sub-processes occur internally in the user. Understanding is achieved when an explanation enables users to clearly perceive its components, mentally process their structure and logic, and derive meaningful insights about the model’s behavior. In this framework, understanding is the result of these coordinated sub-processes, and understandability is the degree to which an explanation facilitates users' success.

\paragraph{Legibility (supports Perception).} Legibility is a property of the explanation that determines how clearly its visual or structural components can be perceived. It affects the user’s ability to visually parse the elements of the explanation. This includes modalities such as heatmaps, textual descriptions, or symbolic rule sets. Legible explanations are characterized by attributes such as contrast, spacing, and form, that render components readily distinguishable.
\begin{itemize}
  \item \textbf{Good Example}: A heatmap with high-contrast colors (e.g., red vs. blue) and sharp boundaries, allowing users to instantly identify influential image regions.
  \item \textbf{Poor Example}: A heatmap with overlapping pastel gradients (e.g., yellow on white), where users struggle to isolate salient features.
\end{itemize}

\paragraph{Readability (supports Comprehension).} Readability is a property of the explanation that reflects how easily its content can be mentally processed once it has been visually accessed. It concerns the clarity and structure of the conveyed information and affects the cognitive effort required to make sense of it. Readable explanations are organized, logically sequenced, and avoid unnecessary complexity, thereby supporting the user’s process of comprehension.
\begin{itemize}
  \item \textbf{Good Example}: A decision tree with concise branching (e.g., $\leq5$ layers) that clearly maps features to outcomes, allowing users to trace logic without cognitive overload. 
  \item \textbf{Poor Example}: A decision tree with $20+$ nested branches, overwhelming users with excessive detail and impeding comprehension.
\end{itemize}

\paragraph{Interpretability (supports Interpretation).} Interpretability is a property of the explanation that reflects the extent to which it supports causal or conceptual reasoning about the model’s inference. It affects the user’s ability to draw meaningful insights, form causal links, and understand why a particular output was produced. Interpretability supports the cognitive process of interpretation, where users integrate information and reason about model behavior.
\begin{itemize}
  \item \textbf{Good Example}: A counterfactual explanation following the structure of \cite{dai2022counterfactual}: "If your income had been \$50k, your loan would have been approved."
  \item \textbf{Poor Example}: A vague statement: "Model output reflects input patterns," offering no causal insight. \newline 
\end{itemize}

\paragraph{Understandability (leads to Understanding).} Understandability is a holistic property of the explanation that reflects its overall capacity to support the user in forming a coherent mental model of the machine learning system’s behavior. It results from the combined effects of legibility, readability, and interpretability, and is influenced by contextual factors such as the display, the goals resulting from task requirements, and the Individual's prior knowledge (see more in section \ref{sec:distasind}). Whilst understandability is an external attribute of the explanation, understanding is the corresponding internal cognitive process --- the mental outcome in the user when perception, comprehension, and interpretation succeed.

\subsection{Influencing Factors: The Display, the Goals and the Individual's Knowledge\label{sec:distasind}}

\begin{table}[hbt!]
    \centering
    \includegraphics[width=1\linewidth]{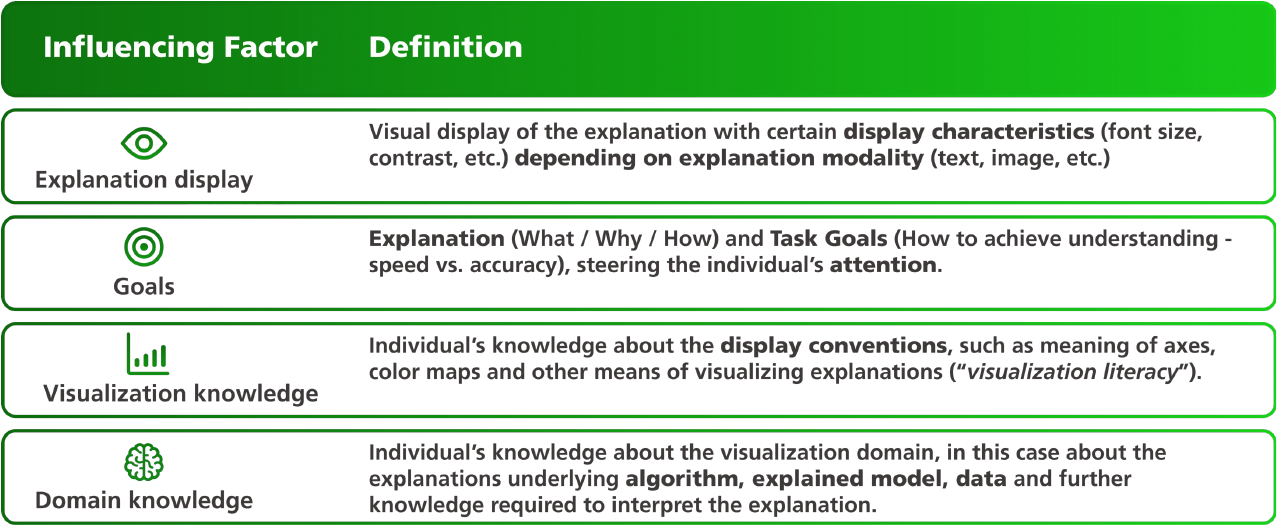}
    \caption{Influencing Factors in the cognitive process and how they affect explanation properties.}
    \label{fig:enter-label}
\end{table}

Factors that influence the cognitive process of understanding explanations can be grouped into three categories: the \textit{display}, the \textit{goals}, and the \textit{individual's knowledge}. These factors interact with the properties of an explanation (i.e., legibility, readability, and interpretability), by shaping how they are perceived, comprehended, and interpreted. As such, they affect the explanation’s ability to support the user’s mental processing and ultimately determine whether it is understandable.

The \textit{display} refers to how the explanation is presented to the user, defined by the chosen modality, whether text, visual, or diagrammatic, and the design decisions that shape the explanation. The display impacts how effectively, easily, or quickly the explanation is perceived, which in turn affects comprehension and interpretation. As such, the display plays a crucial role in the overall understandability of explanations.

\begin{figure}[!ht]
    \centering
    \includegraphics[width=1\linewidth]{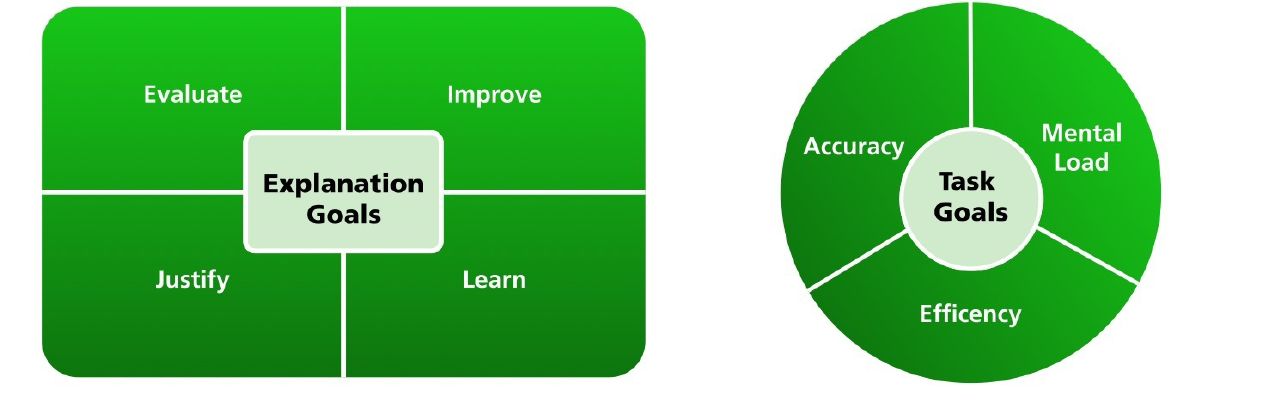}
    \caption{Explanation and task goals of individuals as influencing factors in the cognitive process.}
    \label{fig:xai_goals}
\end{figure}

The task or context provides individuals with \textit{goals} that direct their attention during sensory processing and shape their perception of relevant visual features. When reviewing an explanation, individuals operate within a specific task context, aiming to achieve certain \textit{goals} \cite{narayanan_how_2018}. These objectives can be divided into two categories: \textit{Explanation goals} and \textit{task goals}. Explanation goals refer to the individual's purpose related to the explanation itself, such as evaluating the accuracy of a model's prediction or enhancing its performance \cite{meske2022explainable}. Task goals, on the other hand, define the desired outcomes for task completion, such as minimizing time or effort (see Figure \ref{fig:xai_goals}).

The \textit{individual's knowledge} significantly influences both the comprehension and interpretation of the explanation. Visualization knowledge refers to the individual’s familiarity with display conventions, such as understanding axes, colormaps, and other visualization techniques, often referred to as ``visualization literacy'' \cite{boy2014principled}. Domain knowledge refers to the individual’s expertise in the subject matter of the visualization, including the underlying algorithm, the model being explained, the data, and any additional context needed to interpret the explanation \cite{hegarty_cognitive_2011}.

\subsection{Illustrative example: interpreting heatmaps with the Cognitive Model}
\begin{figure*}[t]
    \centering
    \includegraphics[width=\textwidth]{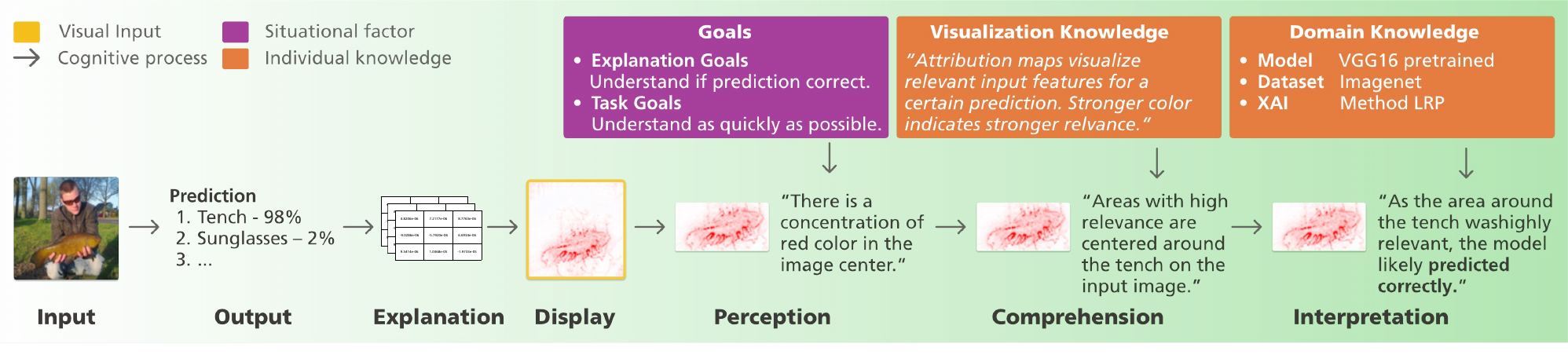}
    \caption{An end-to-end example of applying the cognitive model on the process of understanding an attribution map (\textit{``heatmap''}), computed with LRP for the prediction of the ImageNet class ``Tench''.}
    \label{fig:use_case}
\end{figure*}

An example for better illustration of the cognitive processes in CUE is displayed in Figure \ref{fig:use_case}. When an input image from the ImageNet class ``Tench'' is presented to a classification model, the model outputs a high-likelihood prediction for the class ``Tench''. For explanation, LRP is applied to compute attribution scores for the prediction-relevant input features, which are visualized as an attribution map, or \textit{``heatmap''}. Motivated \textit{explanation} and \textit{task} goals, the individual’s attention is drawn to the most prominent red region located centrally on the display. Utilizing their understanding of heatmaps as a visualization tool, the individual interprets this highlighted region. By integrating domain knowledge of the underlying model, dataset, and explanation method, the individual can reach an interpretation of the explanation, refining their understanding of the model's inference.

\section{Evaluation of the cognitive model}
A central assumption of CUE is that the design of a visual explanation affects the quality of user understanding and decision-making. To evaluate this, we formulated three research questions:
\begin{itemize}
    \item \textbf{RQ1:} How do different colormaps (e.g., BWR, Cividis, Coolwarm) affect the subjective experience (confidence, effort, efficiency) and objective performance of users when interpreting XAI heatmap explanations, particularly for visually impaired individuals?
    \item \textbf{RQ2:} Does the use of accessibility-focused colormaps (e.g., Cividis) mitigate usability gaps between visually impaired and non-impaired users in understanding XAI explanations?
    \item \textbf{RQ3:} How do the cognitive processes outlined in the CUE model align with empirical outcomes when manipulating explanation legibility (via colormaps)?
\end{itemize}

\subsection{Experiment Design}
\begin{figure}[hbt!]
    \centering
    \includegraphics[width=1\linewidth, height=0.45\linewidth]{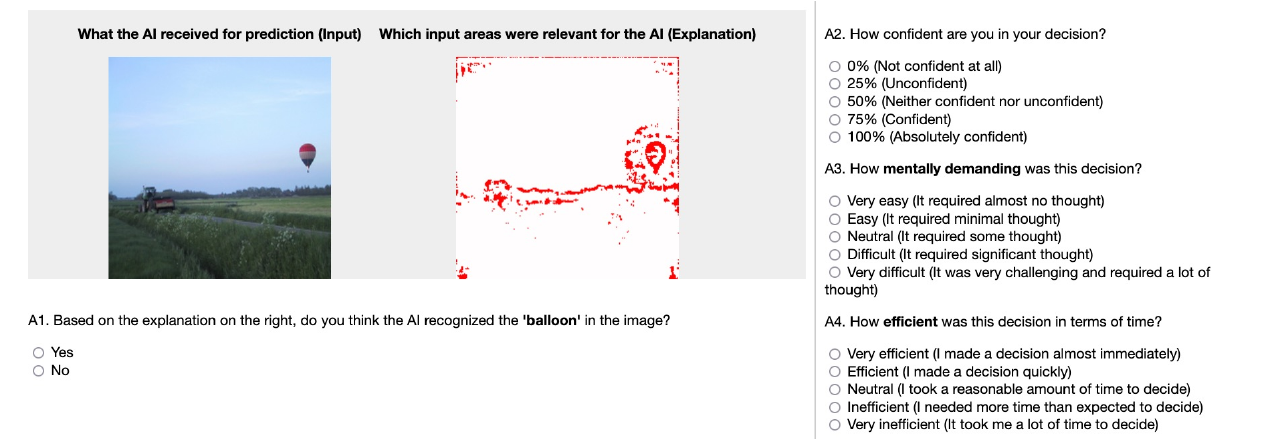}
    \caption{Screenshot of one of the 16 survey questions. This example shows an explanation generated with LRP and displayed with the BWR colormap.}
    \label{fig:question_design}
\end{figure}

We selected image classification with convolutional neural networks (CNN) and heatmap explanations as a proxy task modality as it represents one of the most foundational and widely studied applications in explainable AI. Proxy tasks are controlled experimental setups that simplify real-world explanation scenarios and allow one to isolate variables (e.g., colormap design), control task complexity, and ensure replicability \cite{liao2021human}. In addition, this modality allowed us to control task complexity, standardize stimuli, and ensure replicability, which is essential for an initial investigation.

Three well-examined attribution techniques, namely LRP, Grad-CAM, and LIME, were chosen to represent diverse underlying explanation mechanisms. This choice aims to ensure that any observed effects of colormap are not tied to a single explanation technique. It also aligns with the model's assumption that the form and medium of explanation (that is, the explanation display) interact with cognitive processing.

To address the research questions, we selected three colormaps: BWR, Coolwarm (common but not accessibility-optimized) \cite{matplotlib_colormaps} from Matplotlib \cite{hunter2007matplotlib}, and Cividis, which was specifically designed for accessibility by users with color vision deficiency \cite{nunez2018optimizing}. Participants with and without visual impairments evaluated heatmap explanations across these conditions, completing tasks to measure:
\begin{itemize}
    \item \textbf{Objective performance} (precision, recall, specificity, accuracy, F1 score),
    \item \textbf{Subjective experience} (confidence, effort, efficiency).
\end{itemize}
This design directly operationalizes RQ1 (colormap effects) and RQ2 (accessibility impact). To address RQ3, we aligned each evaluation measure with cognitive processes from CUE. The specific mapping of task questions to these stages is detailed below.

Participants evaluated 16 image-explanation pairs (Figure~\ref{fig:question_design}), and were exposed to one XAI method and all three colormaps. The interface included four questions: binary decision (A1), confidence rating (A2), effort (A3), and efficiency (A4).

Each question was mapped to distinct CUE stages:
\begin{itemize}
    \item \textbf{A1 (Interpretation):} Assesses the user’s ability to judge AI prediction-correctness (interpretability).
    \item \textbf{A2 (Comprehension):} Measures subjective confidence (readability).
    \item \textbf{A3 (Perception):} Captures mental effort (legibility).
    \item \textbf{A4 (Perception + Comprehension):} Reflects processing efficiency (legibility and readability).
\end{itemize}

\subsection{Experiment Execution}

The experiment was executed on the survey platform Amazon Mechanical Turk by Amazon Web Services between 6th to 14th of March 2025. In total, 455 participants were accepted to the study. Of the 455 participants, 78 ($\approx$ 17\%) reported visual impairment, while 377 ($\approx$ 83\%) reported not having impaired vision. Of the 78 participants with self-reported visual impairment, the majority of them reported ``color blindness'' with 55 participants while the remaining 23 reported ``low vision'' or ``other'' visual impairments. Regarding educational background, 306 participants ($\approx$ 67\%) held a bachelor’s degree, 103 ($\approx$ 23\%) held a master’s degree, 40 ($\approx$ 9\%) had school-level education, and 6 ($\approx$ 1\%) held a PhD. Prior experience with XAI varied: 67 ($\approx$ 15\%) participants reported advanced experience, 196 ($\approx$ 43\%) reported an intermediate level, 131 ($\approx$ 29\%) identified as beginners, and 61 ($\approx$ 13\%) had no prior experience with XAI.

\subsection{Methods}

To assess the impact of visual impairment on participants’ performance and experience under different explanation conditions, we first examined for normality using the Shapiro-Wilk test. Normality was assessed in two ways: (1) within each colormap condition for the visually impaired and non-impaired subgroups, and (2) by comparing the differences in outcomes between these subgroups. As the distributions were not normal, we applied Mann-Whitney U-tests for subsequent comparisons. These tests compared groups of participants with and without self-reported visual impairments, each exposed to one of three XAI methods and to all three colormaps. The analysis reports the group size ($n$), $p$-values, and rank-biserial correlation as the effect size ($r$), with positive values indicating an advantage for the group without impairments, and negative values indicating the opposite. The significance threshold for these tests was set at $\alpha = 0.05$.

\section{Results}

\begin{table}
\centering
\begin{tabular}{lcc}
\toprule
\textbf{Metric} & \textbf{Mean} & \textbf{Standard Deviation} \\
\midrule
\multicolumn{3}{l}{\textit{Objective Performance Metrics}} \\
Precision     & 0.521 & 0.113 \\
Recall        & 0.696 & 0.223 \\
Specificity   & 0.332 & 0.254 \\
Accuracy      & 0.514 & 0.102 \\
F1 Score      & 0.575 & 0.123 \\
\midrule
\multicolumn{3}{l}{\textit{Subjective Ratings}} \\
Confidence (\%) & 73.00 & 16.45 \\
Effort (Likert-Scale)          & 3.168 & 0.813 \\
Efficiency (Likert-Scale)        & 3.323 & 0.845 \\
\bottomrule
\end{tabular}
\caption{Average objective and subjective results for all participants.}
\label{tab:avg_results}
\end{table}

\subsection{RQ1: Colormap Effects on Subjective Experience and Objective Performance}

Participants achieved near-chance accuracy (51.4\%, Table~\ref{tab:avg_results}), with no significant differences between colormaps (all $p > 0.05$, Table~\ref{tab:overall_results}). However, subjective experience varied markedly:

\begin{itemize}
    \item Visually impaired users reported lower confidence ($p < 0.01$, $r = 0.304$), higher effort ($p < 0.01$, $r = 0.349$), and reduced efficiency ($p < 0.01$, $r = 0.297$) across all colormaps (Table~\ref{tab:overall_results}).
    \item Coolwarm (non-accessible) exacerbated disparities most severely (efficiency $r = 0.690$), while \textit{Cividis} (accessible) also amplified gaps (efficiency $r = 0.524$). BWR had the least negative impact and was relatively similar compared to the gaps of all groups combined.
\end{itemize}

This dissociation between objective performance (interpretability) and subjective experience (perception/comprehension) suggests that colormaps shape how users engage with explanations, not what they achieve. It may reflect broader challenges related to the proxy task design, the visualization format, or the explanation methods used, especially given that average task accuracy exceeded only slightly chance level while a majority of participants had some degree of experience with XAI. Also, task performance may not only be impacted by factors adjusting the visual presentation, but is also subject to other external explanation qualities such as faithfulness.

\subsection{RQ2: Accessibility-Focused Colormaps and Usability Gaps}

Contrary to expectations, Cividis worsened confidence ($r = 0.462$) and efficiency ($r = 0.524$) disparities compared to BWR ($r = 0.254$), despite its perceptual optimizations. Task performance remained unchanged ($p > 0.05$), suggesting that the presence of ``accessible'' design elements does not necessarily lead to more accessible XAI systems.

\subsection{RQ3: CUE’s Cognitive Stages and Empirical Alignment}

Results validated CUE’s cascading cognitive process:

\begin{itemize}
    \item \textbf{Legibility (Perception):} Effort ($r = 0.349$) aligned with impaired users’ difficulty parsing visual features.
    \item \textbf{Readability (Comprehension):} Lower confidence ($r = 0.304$) reflected struggles to mentally process explanations.
    \item \textbf{Interpretability (Interpretation):} Stable accuracy (51.4\%) showed task outcomes were decoupled from perception/comprehension deficits.
\end{itemize}

The model’s assumption that legibility deficits cascade into downstream cognitive strain was empirically supported --- even if task performance remained unaffected.

\begin{table}[hbt!]
\centering
\begin{tabular}{llccc}
\toprule
\textbf{Colormap} & \textbf{Metric} & \textbf{p-value} & \textbf{Sig.} & \textbf{Effect size} \\
\midrule

\multirow{8}{*}{\shortstack{\textbf{Combined} \\ $n_1=377,$\\ $n_2=78$}} 
  & Precision   & 0.5026 & \ding{55} & -0.0267 \\
  & Recall      & 0.2994 & \ding{55} & -0.0403 \\
  & Specificity & 0.4899 & \ding{55} & -0.0270 \\
  & Accuracy    & 0.2199 & \ding{55} & -0.0484 \\
  & F1          & 0.2500 & \ding{55} & -0.0467 \\
  & \cellcolor{lightgray}Confidence  & \cellcolor{lightgray}0.0000 & \cellcolor{lightgray}\ding{51} & \cellcolor{lightgray}0.3040 \\
  & \cellcolor{lightgray}Effort      & \cellcolor{lightgray}0.0000 & \cellcolor{lightgray}\ding{51} & \cellcolor{lightgray}0.3490 \\
  & \cellcolor{lightgray}Efficiency  & \cellcolor{lightgray}0.0000 & \cellcolor{lightgray}\ding{51} & \cellcolor{lightgray}0.2970 \\
\midrule

\multirow{8}{*}{\shortstack{\textbf{BWR} \\ $n_1=377,$\\ $n_2=78$}} 
  & Precision   & 0.8859 & \ding{55} & -0.0101 \\
  & Recall      & 0.7437 & \ding{55} & -0.0226 \\
  & Specificity & 0.8221 & \ding{55} & 0.0156 \\
  & Accuracy    & 0.8124 & \ding{55} & -0.0166 \\
  & F1          & 0.9794 & \ding{55} & -0.0019 \\
  & \cellcolor{lightgray}Confidence  & \cellcolor{lightgray}0.0000 & \cellcolor{lightgray}\ding{51} & \cellcolor{lightgray}0.2963\hspace{1mm}(↓0.01) \\
  & \cellcolor{lightgray}Effort      & \cellcolor{lightgray}0.0000 & \cellcolor{lightgray}\ding{51} & \cellcolor{lightgray}0.3409\hspace{1mm}(↓0.01) \\
  & \cellcolor{lightgray}Efficiency  & \cellcolor{lightgray}0.0004 & \cellcolor{lightgray}\ding{51} & \cellcolor{lightgray}0.2536\hspace{1mm}(↓0.04) \\
\midrule

\multirow{8}{*}{\shortstack{\textbf{Cividis} \\ $n_1=377,$\\ $n_2=78$}} 
  & Precision   & 0.8827 & \ding{55} & 0.0211 \\
  & Recall      & 0.7405 & \ding{55} & -0.0446 \\
  & Specificity & 0.8810 & \ding{55} & -0.0204 \\
  & Accuracy    & 0.7724 & \ding{55} & -0.0396 \\
  & F1          & 0.9100 & \ding{55} & -0.0167 \\
  & \cellcolor{lightgray}Confidence  & \cellcolor{lightgray}0.0016 & \cellcolor{lightgray}\ding{51} & \cellcolor{lightgray}0.4619\hspace{1mm}(↑0.16) \\
  & \cellcolor{lightgray}Effort      & \cellcolor{lightgray}0.0012 & \cellcolor{lightgray}\ding{51} & \cellcolor{lightgray}0.4743\hspace{1mm}(↑0.13) \\
  & \cellcolor{lightgray}Efficiency  & \cellcolor{lightgray}0.0003 & \cellcolor{lightgray}\ding{51} & \cellcolor{lightgray}0.5238\hspace{1mm}(↑0.23) \\
\midrule

\multirow{8}{*}{\shortstack{\textbf{Coolwarm} \\ $n_1=377,$\\ $n_2=78$}}
  & Precision   & 0.3679 & \ding{55} & -0.1257 \\
  & Recall      & 0.8495 & \ding{55} & -0.0260 \\
  & Specificity & 0.0864 & \ding{55} & -0.2254 \\
  & Accuracy    & 0.1373 & \ding{55} & -0.2031 \\
  & F1          & 0.3804 & \ding{55} & -0.1257 \\
  & \cellcolor{lightgray}Confidence  & \cellcolor{lightgray}0.0004 & \cellcolor{lightgray}\ding{51} & \cellcolor{lightgray}0.5214\hspace{1mm}(↑0.22) \\
  & \cellcolor{lightgray}Effort      & \cellcolor{lightgray}0.0001 & \cellcolor{lightgray}\ding{51} & \cellcolor{lightgray}0.5672\hspace{1mm}(↑0.22) \\
  & \cellcolor{lightgray}Efficiency  & \cellcolor{lightgray}0.0000 & \cellcolor{lightgray}\ding{51} & \cellcolor{lightgray}0.6904\hspace{1mm}(↑0.39) \\
\bottomrule
\end{tabular}
\caption{
Group differences between participants with and without visual impairment across enhancement conditions. 
Significance was tested using the Mann–Whitney U test; effect sizes (rank-biserial correlation) reflect advantages for participants without impairments when positive. 
Arrows indicate changes from the combined group. Colored rows highlight significant differences ($p < 0.05$). 
Results suggest that subjective experiences differ by impairment status and colormap.
}
\label{tab:overall_results}
\end{table}

\section{Limitations}
 Our model makes several assumptions that may not hold universally. It presumes clear and stable user goals, though in practice, users may have evolving or conflicting objectives (e.g., verifying a prediction while also trying to understand model behavior). It also assumes a linear progression from perception to comprehension to interpretation, whereas users may iterate between these stages or apply heuristics (i.e. taking guesses). Moreover, the model is focused on static explanations and may not adequately account for interactive interfaces.

Empirically, our study focused exclusively on colormap manipulations for heatmaps, leaving out other visual attributes such as contrast or sharpness and more advanced forms of explanation visualizations. The tasks were brief and survey-based, which may not fully capture real-world decision-making. Additionally, while we included participants with self-reported visual impairments, the group was heterogeneous, limiting specific subgroup conclusions. Also, the use of only three XAI techniques restricts generalizability across other explanation methods. Finally, we did not disentangle understandability from other effects such as explanation faithfulness from task performance.

\section{Conclusion}

Our study answers three key research questions:

\begin{itemize}
    \item \textbf{RQ1:} Colormaps significantly affect subjective experience (confidence, effort, efficiency) but not objective performance, pointing to a disconnect between user engagement and task outcomes.
    \item \textbf{RQ2:} Accessibility-focused colormaps (in this case Cividis) fail to mitigate usability gaps for visually impaired users, emphasizing the need for holistic designs that address comprehension and interpretation, not just perception.
    \item \textbf{RQ3:} The CUE model’s cognitive stages (\textit{perception} $\rightarrow$ \textit{comprehension} $\rightarrow$ \textit{interpretation}) were empirically validated: legibility deficits cascaded into subjective burdens, even when interpretation (performance) remained stable.
\end{itemize}

These findings challenge the assumption that perceptual optimizations (e.g., accessible colormaps) suffice for inclusive XAI. Instead, they advocate for adaptive interfaces that dynamically adjust explanations to user needs (e.g., multi-modal feedback for impaired users) and user-centered evaluations of ``accessible'' design choices.
Two key insights emerge:
\begin{enumerate}
    \item \textbf{Accessibility by design is not accessibility in practice.} Visually impaired users continued to face greater cognitive effort and lower confidence, even with perceptually uniform colormaps. In some cases, these colormaps worsened their experience.
    \item \textbf{Subjective disadvantages persisted despite equal performance.} Colormap adjustments can impose inequal perceptual and cognitive demands, revealing that visual equity is not guaranteed through perceptual engineering alone.
\end{enumerate}

Also, the findings underscore the need for explanation systems that go beyond visual optimization. Adaptive, multi-modal interfaces, allowing for interactive and tailored engagement, may better support diverse user needs \cite{miller2019explanation,madumal2019grounded,raees2024explainable,labarta2024study}, especially for individuals with limited visual capacity.

Additionally, the observed dissociation between theoretically accessible and less-accessible colormaps highlights the importance of validating design choices in real-world, user-centered contexts. Perceptual models must be complemented by empirical usability testing to ensure actual benefit. Also, the way of empirical testing should be further optimized, as shown by the user performance close to chance.

At a broader level, our results provide initial empirical support for the CUE model: even modest changes in visual presentation (such as colormap choice) influenced how users processed and understood explanations. This suggests that more targeted and sophisticated modifications, e.g., restructured layouts, animation, or interactivity, may offer even greater effects and further validate the model. 

Future work should investigate the interaction between explanation modalities and specific types of visual impairments (e.g., color blindness, low acuity) and explore non-visual formats to reduce reliance on vision altogether.

Combined, these findings advocate for a more inclusive approach to XAI: One that considers not only how explanations are generated and displayed, but how they are actually perceived and understood by diverse users.

\newpage

\section*{Ethical Statement}

The Ethics Commission of the Fraunhofer Heinrich-Hertz-Institute provided guidelines for the study procedure and determined that no protocol approval is required. Informed consent has been obtained from all participants.

\section*{Acknowledgments}

This work was supported by the European Union’s Horizon Europe research and innovation programme (EU Horizon Europe) as grants [ACHILLES (101189689), TEMA (101093003)]; and the German Research Foundation (DFG) as research unit DeSBi [KI-FOR 5363] (459422098).

\bibliographystyle{named}
\bibliography{ijcai25}

\end{document}